\documentclass[%
 reprint,
 amsmath,amssymb,
 aps,
]{revtex4-2}

\usepackage{graphicx}
\usepackage{dcolumn}
\usepackage{bm}
\usepackage{graphicx}
\usepackage{amsmath, amsthm, amssymb}
\usepackage{mathtools}	
\usepackage{dsfont}
\usepackage{float}
\usepackage{xcolor}
\newcommand\unappendix{\par
  \setcounter{apdxsection}{\value{section}}%
  \setcounter{section}{\value{savesection}}%
  \setcounter{subsection}{0}%
  \gdef\thesection{\@arabic\c@section}}
\makeatother

\begin{document}

\preprint{APS/123-QED}

\title{Collective Polaritonic Effects on Chemical Dynamics Suppressed by Disorder}

\author{Juan B. P\'erez-S\'anchez}
\affiliation{Department of Chemistry and Biochemistry, University of California San Diego, La Jolla, CA 92093, USA}
\author{Federico Mellini}
\affiliation{Department of Chemistry and Biochemistry, University of California San Diego, La Jolla, CA 92093, USA}
\author{Noel C. Giebink}
\affiliation{Department of Electrical Engineering and Computer Science, and Physics, University of Michigan, Ann Arbor, Michigan, 48109, USA}
\author{Joel Yuen-Zhou}
\affiliation{Department of Chemistry and Biochemistry, University of California San Diego, La Jolla, CA 92093, USA}
\email{joelyuen@ucsd.edu}

\date{\today}

\begin{abstract}
We present a powerful formalism, disordered collective dynamics using truncated equations (d-CUT-E), to simulate the ultrafast quantum dynamics of molecular polaritons in the collective strong coupling regime, where a disordered ensemble of $N\gg10^{6}$ molecules couples to a cavity mode. Notably, we can capture this dynamics with a cavity hosting a single \textit{effective} molecule with $\sim N_{bins}$ electronic states, where $N_{bins}\ll N$ is the number of bins discretizing the disorder distribution. Using d-CUT-E we conclude that strong coupling,
as evaluated from linear optical spectra, can be a poor proxy for polariton chemistry. For highly disordered ensembles, total reaction yield upon broadband excitation is identical to that outside of the cavity, while narrowband excitation produces distinct reaction yields solely due to differences in the initial states prepared prior to the reaction.
\end{abstract}

\maketitle

\section{Introduction}

Molecular polaritons are quasiparticles formed when molecules are strongly coupled to photons trapped in an optical cavity. These systems exhibit many exotic phenomena attributed to the hybridization of collective material polarization and optical field resonances, and have garnered significant attention as they hold the potential to modify energy transfer mechanisms and enhance chemical reactivity without the need for catalysts \cite{Ebbesen,Hongfei,Blake,Owrutsky,RaphaelReview,Wei,TT,Jorge2,Huo,Herrera,Feist1,GalegoRev,Genes,HuoRev,SKC1,SKC2,TaoLi2}. In the most common scenario, strong light-matter coupling is achieved by a macroscopic number of molecules $N$, where their \textit{collective} interaction with the optical mode exceeds the cavity and molecular linewidths.  
Theoretical methods that can incorporate effects of intermolecular interactions \cite{Spano2,Spano,Zhang}, multiple optical cavity modes \cite{ribeiroNat,Mandal,Cao}, complex vibrational and electronic structure \cite{Vendrellayer,Neepa2,Tichauer,Groenhof,Rubio1,Rubio2,Dominik1,Sukharev,Srihari}, and molecular disorder \cite{Houdre,Schachenmayer3,Borrelli,Zhao,Cohn,Rury,Belyanin,Cao,Cao1,Beljonne,Fassioli}, while considering a large number of molecules, are in much need to explain experimental observations. 

Disorder is an unavoidable feature that can impact polariton transport \cite{Xu,suyabatmaz,allard,Musser}, vibrational dynamics \cite{Schachenmayer3}, superradiance \cite{Chen}, and polaron photoconductivity \cite{Giebink}. In this article, we generalize our recently developed collective dynamics Using truncated equations (CUT-E) method \cite{Juan1}, which allows efficient modeling of the quantum molecular dynamics of an arbitrarily large number of identical molecules, to study disordered ensembles. At the core of CUT-E lies the exploitation of permutational symmetries \cite{Spano,KeelingZeb,KeelingZeb2,KeelingZeb3,Juan1,Cui,Bing,Chen}, which scales the problem from an ensemble of $N\to\infty$ identical molecules down to a single {\it effective} molecule strongly interacting with the cavity mode. This effective molecule differs from an actual single molecule in that its strong coupling to the cavity occurs \textit{solely} at the Franck-Condon (FC) region. We lift the constraint of identical molecules, coarse-grain the disorder distributions, and apply permutational symmetries among molecules that belong to the same disorder ``bin.'' This approach is numerically exact for sufficiently short propagation times, where the number of disorder bins required to reach convergence (denoted as $N_{bins}$) is much smaller than $N$. This seemingly simple strategy has powerful consequences: the resulting d-CUT-E method maps a disordered molecular ensemble into a single effective molecule, with amplified number of electronic degrees of freedom (DoF) by $N_{bins}$. This is both conceptually insightful and computationally efficient compared to conventional methods that include every molecule of the ensemble explicitly. Although our method is general and can be applied to arbitrary disorder distributions of the parameters that shape the potential energy surfaces (PESs), here we focus on Gaussian exciton-frequency disorder and study its impact on various ultrafast molecular and optical polaritonic properties.  

Our study shows that (1) broadband excitation can modify the reaction yield of the ensemble; however, in the large disorder regime, such changes can be explained just by changes in cavity leakage, despite the presence of the polariton bands in the absorption spectrum; and (2) narrowband excitation can modify the reaction yield even in the large disorder regime. In this case, external narrowband laser and strong light-matter coupling allows the selective excitation of high-frequency vibrational states near the upper polariton (UP) band, which are more reactive than the corresponding excitations of the lower polariton (LP) band. Since the polariton processes involved in the preparation of these states vanish before the reaction ensues, this phenomenon should be interpreted as an optical effect and not as a chemical one.


\section{Method}

To illustrate the method, we assume a single-cavity mode, neglect intermolecular interactions, and restrict ourselves to the first excitation manifold. Our molecular model consists of a ground electronic state and two diabatically coupled excited electronic states, where only one of them can couple to the cavity mode. The molecular model is intentionally simplified to uncover the universal photochemical and photophysical behavior of molecular polaritons. In the CUT-E formalism, each disorder bin is represented by a single effective molecule with a collective coupling $g\sqrt{N}\sqrt{P_{i}}$, with $P_{i}$ being the fraction of molecules in the $i$-th bin. Assuming that the molecules are initially in the global ground state, the CUT-E effective Hamiltonian in the large $N$ limit is given by (see Appendix for details)
\begin{align}\label{eq:ham3}
&\hat{H}^{\prime}_{eff}=\nonumber\\
&\omega_{c}|1\rangle\langle 1|+\sum_{i}^{N_{bins}}\left(-\frac{1}{2}\frac{\partial^{2}}{\partial q_{i}^{2}}+V_{e_{1},i}(q_{i})\right)|e_{1,i}\rangle\langle e_{1,i}|\nonumber\\
&+\sum_{i}^{N_{bins}}\left(-\frac{1}{2}\frac{\partial^{2}}{\partial q_{i}^{2}}+V_{e_{2},i}(q_{i})\right)|e_{2,i}\rangle\langle e_{2,i}|\nonumber\\
&+g\sqrt{N}\sum_{i}^{N_{bins}}\sqrt{P_{i}}\left(|e_{1,i}\rangle\langle 1|+|1\rangle\langle e_{1,i}|\right)\mathds{P}_{i}\nonumber\\
&+v_{12,i}|e_{1,i}\rangle\langle e_{2,i}|+\text{H.c.}.
\end{align} Here, $q_{i}$ is the vector of mass-weighted coordinates of all vibrational degrees of freedom of molecule $i$; $|g_{i}\rangle$, $|e_{1,i}\rangle$, and $|e_{2,i}\rangle$ are the ground and excited electronic states; $V_{g/e_{1}/e_{2}}(q)$ are the PESs; $v_{12,i}$ is the diabatic coupling; $\omega_{c}$ is the cavity frequency; $\hat{a}$ is the photon annihilation operator; and $\mathds{P}_{i}=|\varphi_{0,i}\rangle\langle \varphi_{0,i}|$ is the projector onto the FC state of the $i$-th effective molecule. Notice that the cavity mode only couples to the transition to the $|e_{1}\rangle$ electronic state of the molecules. This Hamiltonian incorporates the collective interaction between the molecules and the cavity mode, neglects all single-molecule coupling effects (which is valid for ultrafast processes if single-molecule coupling strength is sufficiently weak), and neglect finite temperature effects \cite{Juan1}. 

Although the CUT-E Hamiltonian is much simpler than conventional Hamiltonians granted $N_{bins}\ll N$, it still involves the vibrational and electronic degrees of freedom of all bins. Notably, $\hat{H}^{\prime}_{eff}$ can be simplified even more by mapping it to another Hamiltonian $\hat{H}_{eff}$ [Eq. (\ref{eq:ham4})] whose Hilbert space increases \textit{linearly} with $N_{bins}$ (see the Appendix). This linear scaling is a consequence of the following considerations: first, restriction to the first excitation manifold implies that only one of the molecules is electronically excited at a time. Second, molecules in the ground electronic state do not exhibit vibrational dynamics, since the only mechanism by which they can acquire phonons is emission away from the FC region, which can be neglected in ultrafast dynamics according to CUT-E \cite{KeelingZeb,Juan1}. These two features imply that the exact wavefunction of the system is a superposition of states in which only one effective molecule showcases vibrational dynamics, while the other $N_{bins}-1$ are frozen. 

\begin{figure*}[htb]
\begin{center}
\includegraphics[width=1\linewidth]{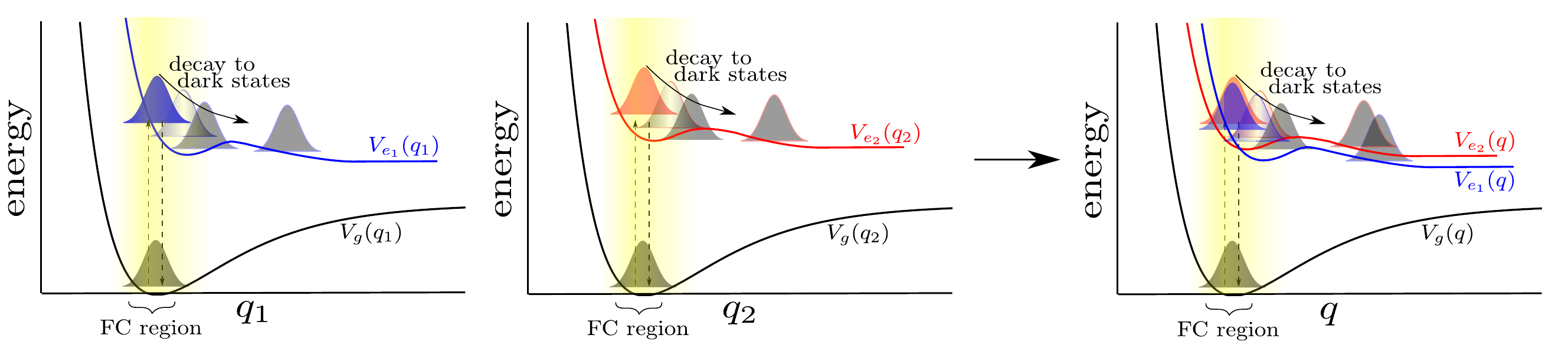}
\caption{Linear scaling of the dynamics with the number of disorder bins. The first two panels depict cuts of the PESs along the reaction coordinate of two different bins, $q_{1}$ and $q_{2}$. d-CUT-E projects the two-dimensional vibrational dynamics of the two bins onto a single nuclear degree of freedom $q$ (third panel).}
\label{fig:linear}
\end{center}
\end{figure*}

We assume disorder only affects the excited state PESs $V_{e,1}(q)$ and $V_{e,2}(q)$. In this way, disorder bins only appear as additional ``electronic'' states. This dramatic reduction of the original system into a single effective molecule showcasing $2N_{bins}$ excited electronic states is the main computational contribution of this manuscript. Its linear scaling with disorder represents a considerable improvement over methods that add an increasing number of molecules with parameters sampled from a disorder distribution, a computationally costly exercise that scales \textit{exponentially} with $N$. A schematic representation of d-CUT-E is shown in Fig. \ref{fig:linear} using two disorder bins as an example, and the mathematical procedure is explained in detail in the Appendix. 

The d-CUT-E Hamiltonian reads
\begin{align}\label{eq:ham4}
\hat{H}_{eff}&=\omega_{c}|1\rangle\langle 1|+\sum_{i}^{N_{bins}}\left(-\frac{1}{2}\frac{\partial^{2}}{\partial q^{2}}+V_{e_{1,i}}(q)\right)|e_{1,i}\rangle\langle e_{1,i}|\nonumber\\
&+\sum_{i}^{N_{bins}}\left(-\frac{1}{2}\frac{\partial^{2}}{\partial q^{2}}+V_{e_{2,i}}(q)\right)|e_{2,i}\rangle\langle e_{2,i}|\nonumber\\
&+g\sqrt{N}\sum_{i}^{N_{bins}}\sqrt{P_{i}}\left(|e_{1,i}\rangle\langle 1|+|1\rangle\langle e_{1,i}|\right)\mathds{P}\nonumber\\
&+\sum_{i}^{N_{bins}}v_{12,i}|e_{1,i}\rangle\langle e_{2,i}|+\text{H.c.}
\end{align} 
Here, $\mathds{P}=|\varphi_{0}\rangle\langle\varphi_{0} |$. The values for $P_{i}$ and the parameters that define the PESs can be obtained from a disorder distribution $\rho(\omega)$. 

\begin{figure}[htb]
\begin{center}
\includegraphics[width=1\linewidth]{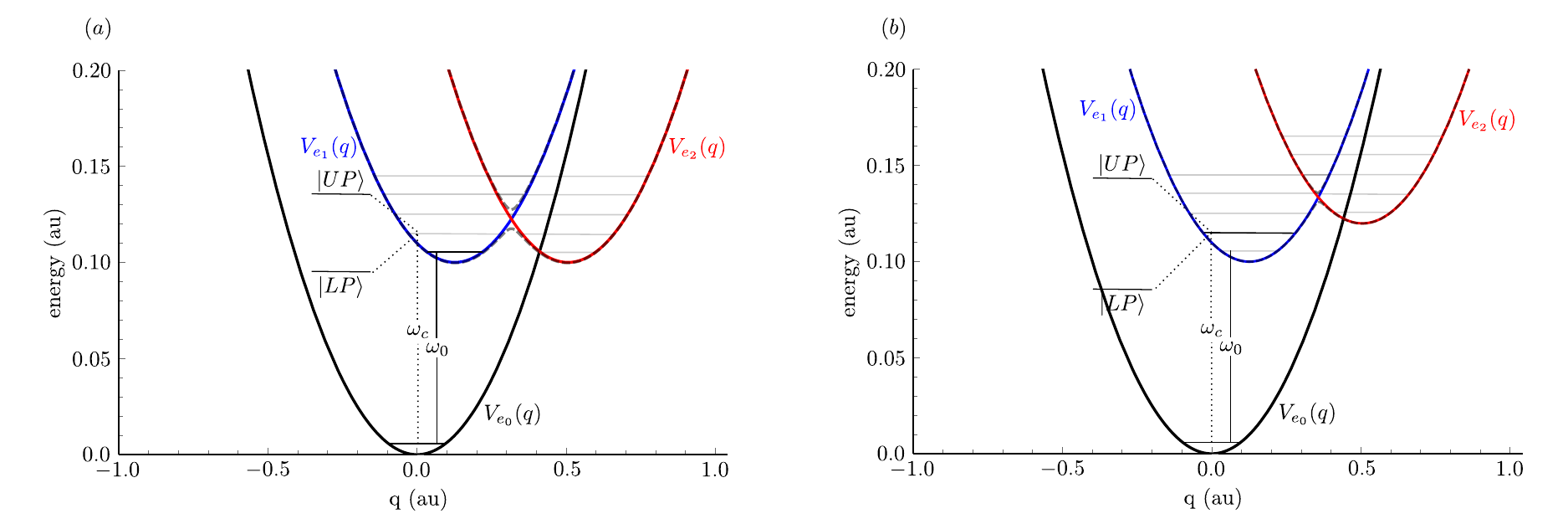}
\caption{Potential energy curves for the molecular system. Each molecule in our model consists of two diabatically-coupled excited electronic states and one vibrational degree of freedom.}
\label{fig:PESs}
\end{center}
\end{figure}

As a proof of principle, and since we are interested in short-time processes that can be modified by strong-light matter interaction, we assume that short-time vibrational dynamics occurs only along the reaction coordinate and ignore vibrational degrees of freedom orthogonal to it \cite{Dvira,CederbaumPRL}. Hence, we use a single vibrational coordinate per molecule (see Fig. \ref{fig:PESs}). 

The final Hamiltonian yields
\begin{align}\label{eq:ham5}
\hat{H}_{eff}&=\left(\omega_{c}-i\kappa/2\right)|1\rangle\langle 1|\nonumber\\
&+\sum_{i}^{N_{bins}}\left(\omega_{0,i}+\omega_{\nu,i}\hat{D}(s_{1,i})\hat{b}^{\dagger}\hat{b}\hat{D}^{\dagger}(s_{1,i})\right)\otimes|e_{1,i}\rangle\langle e_{1,i}|\nonumber\\
&+\sum_{i}^{N_{bins}}\left(\omega_{0,i}+\omega_{\nu,i}\hat{D}(s_{2,i})\hat{b}^{\dagger}\hat{b}\hat{D}^{\dagger}(s_{2,i})\right)\otimes|e_{2,i}\rangle\langle e_{2,i}|\nonumber\\
&+g\sqrt{N}\sum_{i}^{N_{bins}}\sqrt{P_{i}}\left(|e_{1,i}\rangle\langle 1|+|1\rangle\langle e_{1,i}|\right)\mathds{P}\nonumber\\
&+\sum_{i}^{N_{bins}}v_{12,i}|e_{1,i}\rangle\langle e_{2,i}|+\text{H.c.},
\end{align} where $\hat{D}(s_{1,i})=e^{\left(\hat{b}^{\dagger}-\hat{b}\right)s_{1,i}}$ is the displacement operator, and $\omega_{0,i}$, $\omega_{\nu,i}$, and $s_{1,i}$ are the exciton frequency, vibrational frequency, and Huang–Rhys factor for the $|e_{1,i}\rangle$ electronic state respectively. We incorporate cavity leakage by adding the imaginary term $i\kappa/2$ to the photon frequency, and the amount of energy that escapes the cavity by this mechanism can be quantified by calculating the norm of the wave function $\langle\Psi(t)|\Psi(t)\rangle$. For the rest of this work we will consider only exciton-frequency disorder. Here, disorder affects both $|e_{1}\rangle$ and $|e_{2}\rangle$ equivalently, thus the height of the barrier is the same for all bins.

To study optical effects, we calculate the linear absorption spectrum \cite{Cwik,KeelingZeb,HerreraPRA,Arghadip},
\begin{equation}
    A(\omega)=\kappa\operatorname{Re}\left[\tilde{C}(\omega)\right]-\frac{1}{2}\kappa^{2}|\tilde{C}(\omega)|^{2},
\end{equation} with $\tilde{C}(\omega)=\int_{0}^{T_{f}}dt e^{i\omega t}\langle\Psi(0)|\Psi(t)\rangle$ and $| \Psi(0) \rangle=|\varphi_{0}\rangle\otimes|1\rangle$. 
To study chemical effects, we calculate populations of the electronic states $|e_{1}\rangle$ and $|e_{2}\rangle$, which can be thought of as the ``reactant" and ``product" states of a photochemical reaction, respectively. The reaction proceeds via tunneling of the vibrational wave packet from the $|e_{1}\rangle$ state. We calculate populations for each bin, and add them up to obtain the total excited state population of reactant and product in the ensemble,
\begin{align}
    &P_{e_{1}}(t)=\sum_{i}^{N_{bins}}P_{e_{1,i}}(t)=\sum_{i}^{N_{bins}}\langle\Psi(t)|e_{1,i}\rangle\langle e_{1,i}|\Psi(t)\rangle,\nonumber\\
    &P_{e_{2}}(t)=\sum_{i}^{N_{bins}}P_{e_{2,i}}(t)=\sum_{i}^{N_{bins}}\langle\Psi(t)|e_{2,i}\rangle\langle e_{2,i}|\Psi(t)\rangle.
\end{align} To capture the competition between cavity leakage and excited-state reactivity, we do not divide the expectation values by the norm.

\begin{figure}[htb]
\begin{center}
\includegraphics[width=1\linewidth]{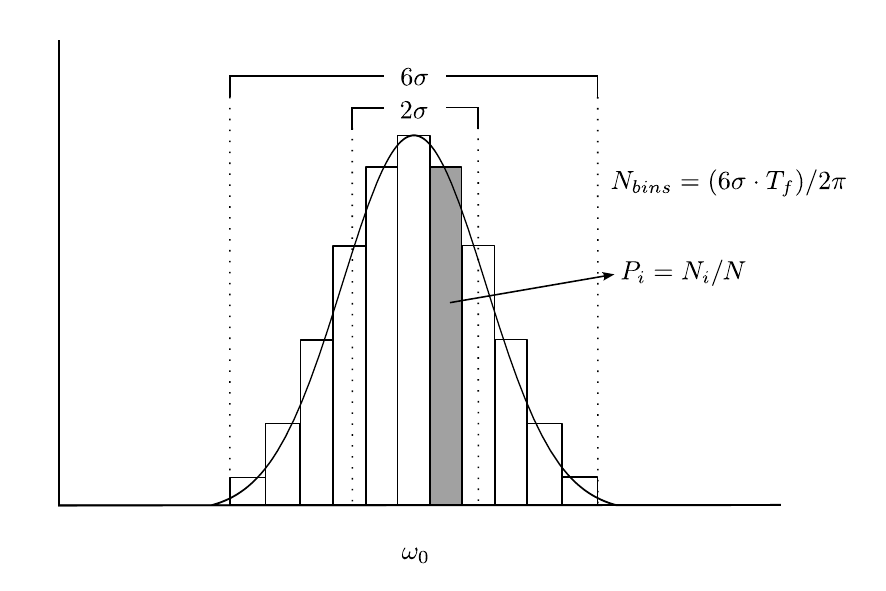}
\caption{Discretization of exciton frequency disorder. The frequencies follow a Gaussian distribution and the number of bins required to reach convergence is proportional to the total disorder and final propagation time.}
\label{fig:disorder}
\end{center}
\end{figure}

Both optical and molecular properties are calculated for various values of exciton-frequency disorder $\sigma$ and collective light-matter coupling $g\sqrt{N}$. We use the Gaussian exciton-frequency distribution $\rho(\omega)=\frac{1}{\sigma\sqrt{2\pi}}e^{-\frac{1}{2}\left(\frac{\omega-\omega_{0}}{\sigma}\right)^2}$, with the cavity frequency being resonant with the $\nu=0\rightarrow\nu^{\prime}=1$ transition $\omega_{c}=\omega_{0}+\omega_{\nu}$. We  discretize $\rho(\omega)$ by restricting its domain to $\omega_{0}-3\sigma<\omega<\omega_{0}+3\sigma$, and calculate the values of $P_{i}$ and $\omega_{0,i}$ as
\begin{equation}    P_{i}=\int_{\omega_{i,min}}^{\omega_{i,max}}d\omega\rho(\omega)\ \ \ \ \ \omega_{0,i}=\int_{\omega_{i,min}}^{\omega_{i,max}}d\omega\omega\rho(\omega),
\end{equation} with $\omega_{i,min}=\omega_{0}-3\sigma+(i-1)(6\sigma/N_{bins})$ and $\omega_{i,max}=\omega_{0}-3\sigma+i(6\sigma/N_{bins})$. Ignoring the tails of the distribution is justified since molecules that fall at the ends of the energy distribution are off-resonant with the cavity mode and the polariton windows (see Fig. \ref{fig:disorder}).

We find that the width of the bins $\delta\omega$ required to reach convergence in optical and chemical observables obey a simple relation $\delta \omega\sim 2\pi/T_{f}$, where $T_{f}$ is the final propagation time. This is not surprising since a finite propagation time implies a finite energy resolution. Thus, the number of bins required for convergence obeys $N_{bins}=6\sigma T_{f}/2\pi$ (see the Appendix B for convergence analysis). Hence, the computational cost of disorder does not scale with $N$, and scales linearly with $N_{bins}\propto T_{f}$.

\section{Results}

\begin{figure*}[htb]
\begin{center}
\includegraphics[width=1\linewidth]{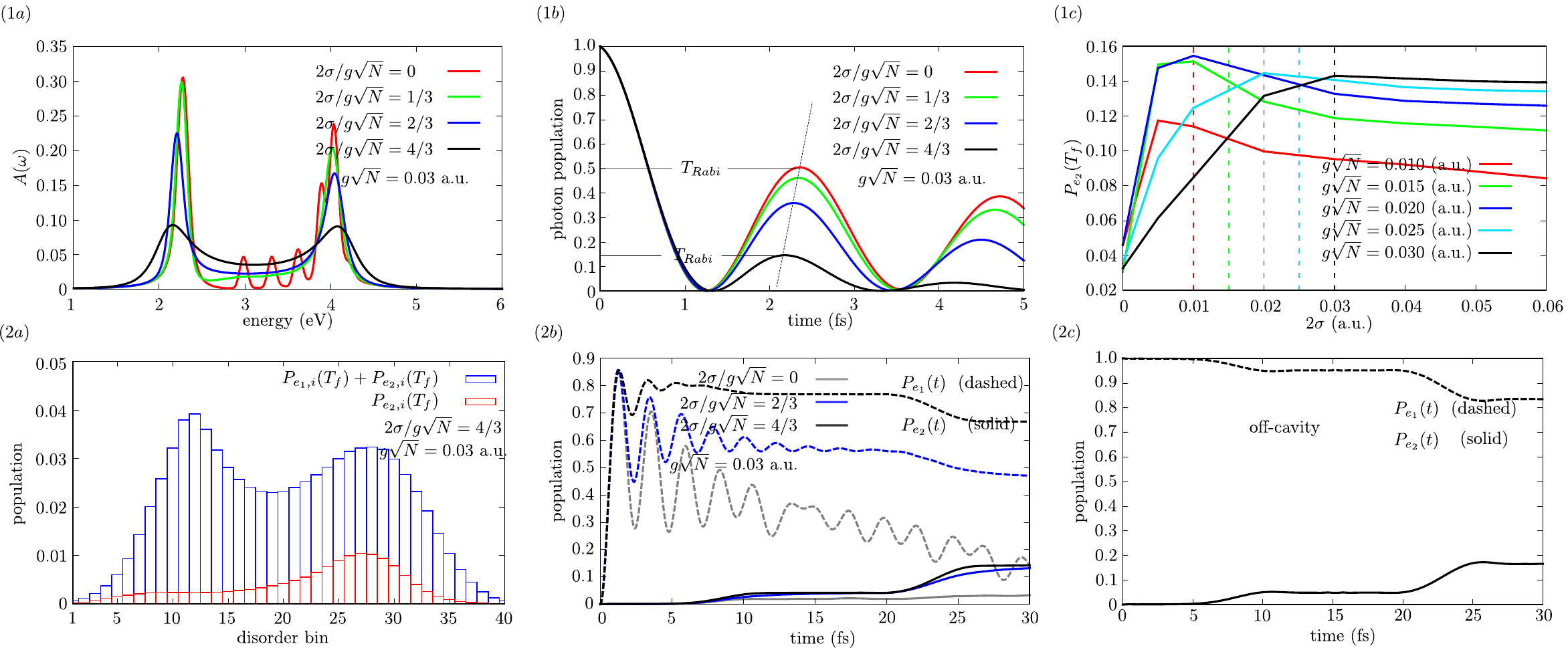}
\caption{Effects of exciton-frequency disorder on optical and chemical properties of molecular polaritons. Parameters: $\omega_{0}=0.10$ au, $\omega_{c}=0.11$ au, $\omega_{\nu}=0.01$ au, $\kappa=0.006$ au, $v_{12}=0.0025$ au, $s_{1}=-1$, and $s_{2}=-4$. (1a) Linear absorption spectrum for different values of disorder. Disorder suppresses vibronic features and increases the polariton Rabi splitting. (1b) Time-dependent interpretation of Rabi splitting increase caused by disorder. (1c) Final population of the electronic state $|e_{2}\rangle$ [$P_{e_{2}}(T_{f})$, $T_{f}=30$ fs] for different values of disorder and collective light-matter coupling strength. Vertical lines on the disorder axis correspond to $2\sigma=g\sqrt{N}$. As expected, polaritonic effects become immune to disorder for large $g\sqrt{N}$.  (2a) Final excited-state populations of each disorder bin show their reaction yield (red) is different even at large disorder, and not just due to polariton modified absorption (blue). Bins are ordered from low to high exciton frequency. (2b) Population dynamics of excited electronic states $P_{e_{1}}(t)$ (dashed) and $P_{e_{2}}(t)$ (solid) for $2\sigma=0$ (gray), $2\sigma<g\sqrt{N}$ (blue), and $2\sigma>g\sqrt{N}$ (black). (2c) Population dynamics of excited electronic states $P_{e_{1}}(t)$ (dashed) and $P_{e_{2}}(t)$ (solid) outside of the cavity. The step-like behavior in $P_{e_{2}}(t)$ is attributed to the reaction proceeding via tunneling of the oscillating vibrational wavepacket from the $|e_{1}\rangle$ state. Notice the resemblance between the dynamics in (2c) and (2b) for $2\sigma/g\sqrt{N}=2/3,4/3$, indicating no polaritonic changes in the PESs in the presence of disorder.}
\label{fig:results}
\end{center}
\end{figure*}

\textit{Broadband excitation}.--- Figure \ref{fig:results} shows optical and chemical effects of disorder for $T_{f}=30$ fs, $N_{bins}=40$, and an initially photonic wavepacket $| \Psi(0) \rangle=|\varphi_{0}\rangle\otimes|1\rangle$ (mimicking broadband excitation, $T_{pulse}\ll T_{Rabi}$). Final propagation time was chosen to avoid barrier recrossing due to the single-mode nature of our model. 

Our calculations show that features that would be resolved at long timescales, such as vibronic progressions (small red peaks), vanish as disorder grows comparable to the collective light-matter coupling [Fig. \ref{fig:results}(1a)]. This is a consequence of damping of the coherent return of population between the bright state to the cavity mode [see Fig. \ref{fig:results}(1b)]. However, contrary to what intuition suggests, even if such dampening is significant within the timescale of the Rabi period $T_{Rabi}\sim 2.5$ fs, it does not imply that the UP and LP bands disappear. This is because only a small amplitude needs to return to the cavity (during the second half of the Rabi cycle) to create such an optical feature \cite{Heller,Tannor}. The reduction in $T_{Rabi}$ that follows from the dampening of Rabi oscillations also leads to an increase in the Rabi splitting \cite{Gera,Cohn}. Thus, for highly disordered polaritons, collective light-matter coupling strengths \textit{cannot} be directly extracted from polariton Rabi splitting. 

Figure \ref{fig:results}(1c) shows changes in $P_{e_{2}}(T_{f})$ that correspond to two regimes of disorder. The low disorder regime $2\sigma<g\sqrt{N}$, characterized by a strong dependence of the reaction yield with $\sigma$, and the large disorder regime $2\sigma>g\sqrt{N}$, where the reaction yield reaches a constant value. For low $\sigma$ and low $g\sqrt{N}$ (red, green and blue lines), disorder leads to higher total reaction yields, but this effect vanishes for large $g\sqrt{N}$ (light blue and black lines) where ``polaron decoupling'' takes over \cite{HerreraPRL,Feist1}. This behavior of the reaction yield in the low disorder regime arises as a consequence of interferences between the vibrational wavepackets of the electronic states for each bin $|e_{1,i}\rangle$, due to their common interaction with the photon state $|1\rangle$. These Rabi oscillations are significantly reduced at large disorder, causing the total reactivity of the ensemble to be independent of disorder. In this regime, changes of the total reaction yield upon broadband excitation for different collective coupling strengths can be explained by differences in cavity leakage (see normalized product populations in Appendix C).

Figure \ref{fig:results}(2a) shows that wave packet interferences in the large-disorder regime still cause differing reactivities on \textit{individual} bins, despite them having the same PESs. By comparing the total excited state population $P_{e_{1}}(t)+P_{e_{2}}(t)$ with the product population $P_{e_{2}}(t)$, we conclude that this difference in product yields of individual bins cannot be attributed to differences in their respective absorption. Additional calculations show that high-frequency bins become more reactive than those at low frequency as a consequence of vibronic coupling of the $|e_{1}\rangle$ electronic state (here, $s_{1}=-1$), and that this effect is suppressed if $s_{1}=0$ (see Appendix D). We will elaborate on the consequences of this effect when we consider narrowband excitation. Figure \ref{fig:results}(2b) shows time-dependent total population dynamics for zero, intermediate, and large disorder. In the absence of disorder, strong light-matter coupling gives rise to large amplitude Rabi oscillations for $P_{e_{1}}(t)$ that last even after the reaction occurs, producing a low reaction yield. This polaritonic-induced reduction of reactivity (polaron decoupling) has been explained in previous theoretical works as a change in the PESs that prevents the nuclei from moving away from the FC region\cite{HerreraPRL,Feist1,Cui,Kuttruff}. However, for intermediate and large disorder, Rabi oscillations are damped before the reaction ensues, polaron decoupling disappears, and the reactivity reaches a constant value. This constant value corresponds to the reaction yield outside of the cavity, as shown in Fig. \ref{fig:results}(2c), where we have initiated the system in the bright state $|\Psi(0)\rangle=\sum_{i}^{N_{bins}}\sqrt{P_{i}}|\varphi_{0}\rangle\otimes|e_{i}\rangle$, for $g\sqrt{N}=0$. From this we conclude that changes in the reaction yield upon broadband excitation are not possible if reactivity occurs on a timescale longer than that of Rabi oscillations, even if the linear absorption spectrum showcases two clear polariton bands [see Fig. \ref{fig:results}(a)].

Next, we systematically study how the large-disorder regime approaches the off-cavity limit by initiating the dynamics in the bright state, and calculating the total excited states populations for $g\sqrt{N}=0.03$ au, increasing the disorder. This allows us to directly compare the ensuing dynamics with that outside of the cavity ($g\sqrt{N}=0$). Importantly, the out-of-cavity population dynamics is independent of disorder for these initial conditions.

\begin{figure*}[htb]
\begin{center}
\includegraphics[width=1\linewidth]{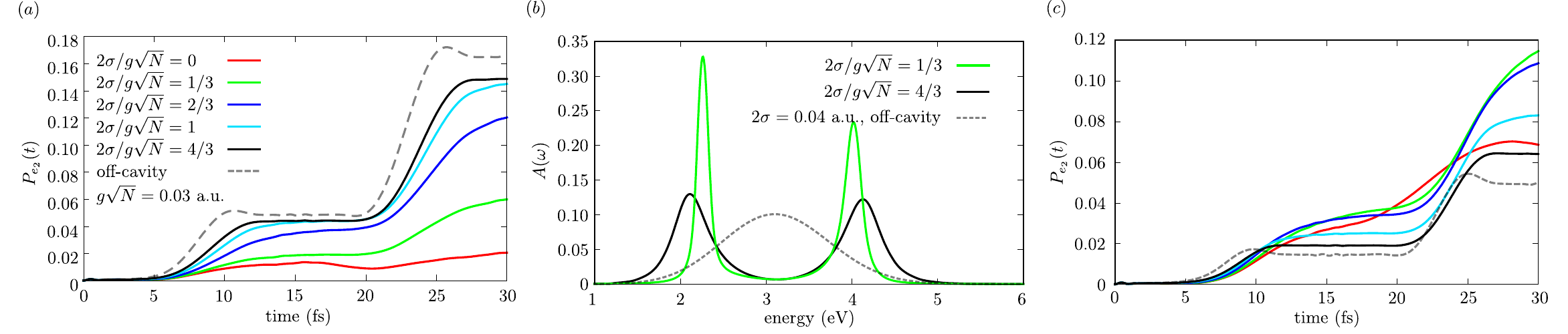}
\caption{Reaction yield in the strong-coupling regime as a function of disorder upon broadband excitation (bright initial state). (a) Reaction is suppressed with strong coupling; however, as disorder becomes comparable to the Rabi splitting, $P_{e_{2}}(t)$ resembles the behavior outside of the cavity. (b) Absorption spectrum $A(\omega)$ for the dynamics in (a): strong coupling low disorder (green), strong coupling-large disorder (black), and outside of the cavity (dashed). Notably, despite well-defined polariton bands for large disorder, the corresponding reactivity in (a) is similar to the off-cavity case. (c) Same as (a) but shifting the PES of $|e_{2}\rangle$ upwards by two vibrational quanta. In this case, strong coupling enhances the reaction for weak disorder.}
\label{fig:results_b}
\end{center}
\end{figure*}

We find that broadband excitation in the large-disorder regime does not result in different photochemistry than outside of the cavity, although the linear absorption spectrum misleadingly suggests otherwise. As we discussed before, this occurs because, even for large disorder, the Rabi splitting in the absorption spectrum is a feature defined at short times ($T\sim 2.5$ fs), for which disorder has a very minor effect. On the contrary, at the longer timescales of chemical reactivity, disorder has already caused decoherence of wavefunction in individual disorder bins. We also modified the PES of $|e_{2}\rangle$ so that it lies at higher energies with respect to $|e_{1}\rangle$. Contrary to the previous scenario, for low disorder, we observe a polariton-mediated increase in the total reaction yield if the target state $|e_{2}\rangle$ is $0.02$ au blue detuned with respect to $|e_{1}\rangle$, thanks to the UP having high enough energy to allow the barrier crossing. Moreover, we observe that the behavior is not monotonic and population transfer is enhanced for low values of disorder, but eventually converges to the dynamics outside of the cavity [see Fig. \ref{fig:results_b}(c)].

\textit{Narrowband excitation}.--- We next examine selective excitation of either one of the polariton bands (narrowband excitation, $T_{pulse}>T_{Rabi}$). To this end, we initiate the dynamics in the states $|\Psi_{\pm}(0)\rangle=\frac{1}{\sqrt{2}}|\varphi_{0}\rangle\otimes\left(|1\rangle\pm\sum_{i}^{N_{bins}}\sqrt{P_{i}}|e_{i}\rangle\right)$, which allocate the same vibrational energy to all bins, and their average energy is that of the polaritons.

\begin{figure*}[htb]
\begin{center}
\includegraphics[width=1\linewidth]{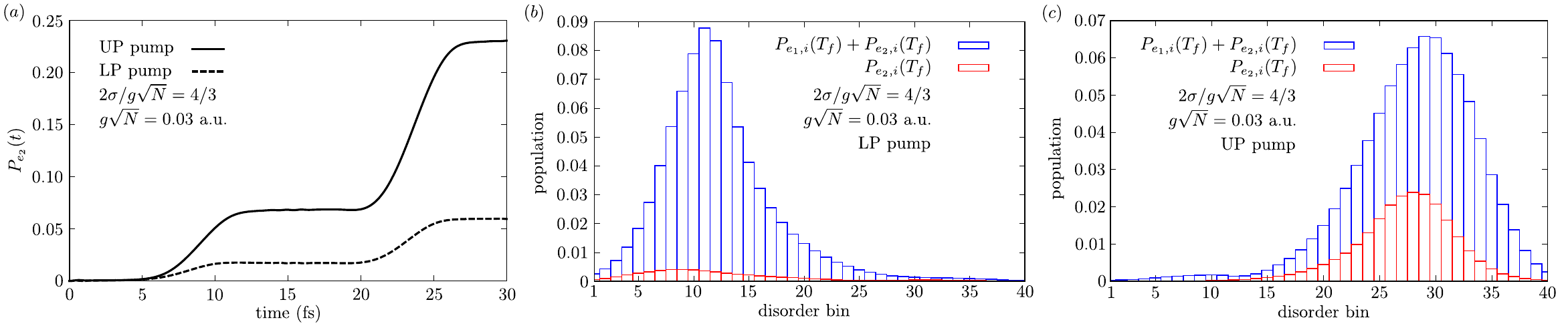}
\caption{Reaction yield upon narrowband excitation in the strong-coupling and large-disorder regime (parameters: $g\sqrt{N}=0.03$ au and $2\sigma=0.04$ au). (a) Total product state population after pumping the UP (solid) and the LP (dashed). (b and c) Excited-state populations of each disorder bin showing their reaction yield (red) and absorption (blue), after LP pumping  and UP pumping. Larger ratios $P_{e_{2,i}}(T_{f})/[P_{e_{1,i}}(T_{f})+P_{e_{2,i}}(T_{f})]$ imply higher reactivity. (b) and (c) indicate that the higher reactivity of the UP bins cannot be explained simply by polariton absorption, which slightly favors the LP bins.}
\label{fig:results1e}
\end{center}
\end{figure*}

Figure \ref{fig:results1e} shows that selective pumping of polariton bands yields different excited-state reaction yields. As we might expect from Fig. \ref{fig:results}(2a), excitation of the UP results in a high reaction yield due to selective excitation of the bins with a higher reactivity, while excitation of the LP results in a low reaction yield. This can be explained as a cavity-assisted mixing of high-lying vibronic states of the low-energy bins with low-lying vibronic states of the high-frequency bins through Rabi oscillations, resulting in excitation of molecules near the UP with more vibrational energy, and molecules near the LP with lower vibrational energy. A scheme of this mechanism is shown in Fig. \ref{fig:model3}, and numerical evidence is provided in Appendix D. Notice that this mechanism is a consequence of the definition of the initial excited state. Since the laser pulse must be longer than the Rabi period to selectively excite a single polariton band, both the external laser and polariton dynamics simultaneously participate in the creation of the initial states with differing reactivities. Crucially, there is no reason to believe that these highly reactive states cannot be created without the cavity's presence using a \textit{different} narrowband linear external laser that efficiently targets high-energy vibronic progressions through the UP (LP) band. In fact, both strategies rely on non zero vibronic coupling to produce frequency-dependent photoreactivity. Therefore, whether collective strong coupling in the large $N$ limit provides any advantage to control chemical reactivity compared to conventional linear optical sources is not readily apparent, especially in light of recent experiments that suggest there are no polaritonic effects on chemical dynamics \cite{thomas,Groenhoff}.

\begin{figure*}[htb]
\begin{center}
\includegraphics[width=1\linewidth]{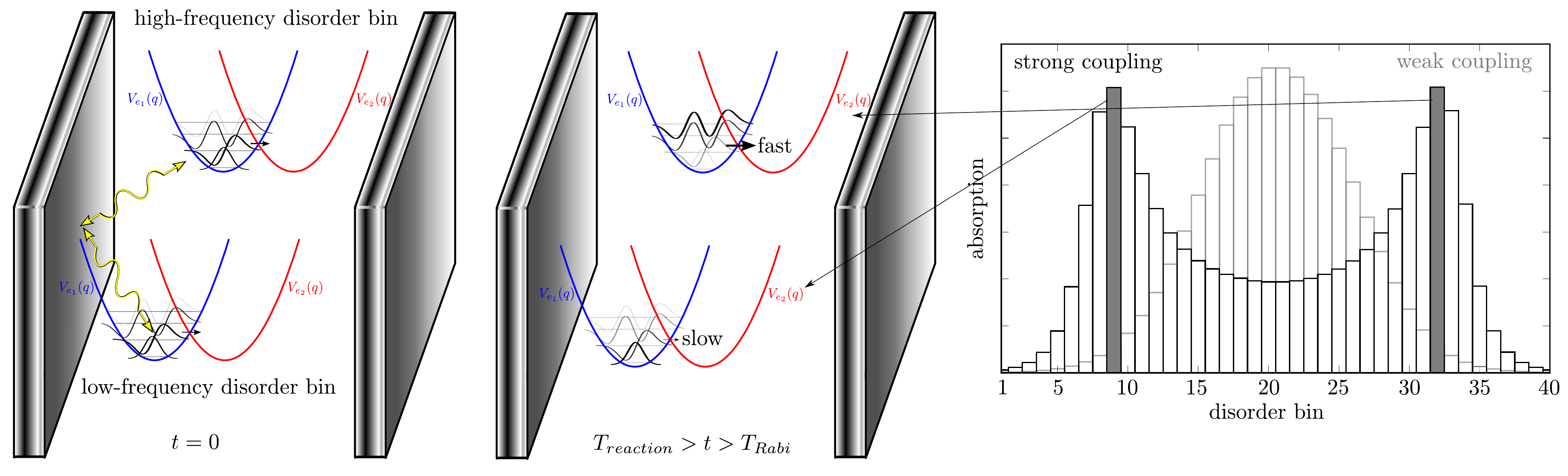}
\caption{Mechanism of frequency-dependent photoreactivity in disordered molecular polaritons. Interplay of narrowband excitation and cavity-mediated interactions between disorder bins results in the preparation of states with different reactivities after damping of Rabi oscillations and before the chemical reaction. These highly (slightly) reactive bins can be excited at the UP (LP) frequency, leading to an increase (decrease) of the total reactivity upon narrowband excitation.}
\label{fig:model3}
\end{center}
\end{figure*}

\section{Summary}

We have successfully introduced the d-CUT-E method to efficiently simulate the ultrafast dynamics of disordered molecular ensembles under collective strong coupling. Our findings challenge the notion that a Rabi splitting in the linear absorption spectrum inevitably implies changes in chemical reactivity. This is due to the disparity in timescales governing optical and chemical properties. Optical properties primarily manifest at short timescales when disorder has minimal impact, while chemical properties emerge at longer timescales. In the common scenario that disorder is comparable with the light-matter coupling strength, strong coupling induces modifications in the reactivity of individual disorder bins, which can be selectively targeted by narrowband pulses. This phenomenon should be interpreted as a cavity-mediated initial state preparation effect.\\

\begin{acknowledgments}
This work was supported by the Air Force Office of Scientific Research (AFOSR) through the Multi-University Research Initiative (MURI) Program No. FA9550-22-1-0317. We thank Gerrit Groenhof, Arghadip Koner, and Kai Schwennicke for useful discussions.
\end{acknowledgments}

\appendix

\section{CUT-E Hamiltonian in the large $N$ limit for discrete disorder}

For simplicity, let us derive $\hat{H}_{eff}$ when there is a single electronic excited state per molecule. The general polariton Hamiltonian yields

\begin{equation}\label{eq:original}
\hat{H}=\sum_{i}^{N}\left(\hat{H}_{mol}^{(i)}+\hat{H}^{(i)}_{I}\right)+\hat{H}_{cav},
\end{equation} where the Hamiltonians $\hat{H}_{mol}^{(i)}$, $\hat{H}_{cav}$, and $\hat{H}^{(i)}_{I}$ describe the energy of the $i$-th molecule, the cavity mode, and the light-matter interaction. In our work we use a simple model that assumes a single cavity mode and neglects intermolecular interactions, leading to

\begin{align}
\hat{H}^{(i)}_{mol}&=\hat{h}^{(i)}_{g}|g_{i}\rangle\langle g_{i}|+\hat{h}^{(i)}_{e}|e_{i}\rangle\langle e_{i}|,\nonumber\\
&=-\frac{1}{2}\frac{\partial^{2}}{\partial q_{i}^{2}}\mathds{1}+V_{g,i}(q_{i})|g_{i}\rangle\langle g_{i}|+V_{e,i}(q_{i})|e_{i}\rangle\langle e_{i}|,\nonumber\\
\hat{H}_{cav}&=\omega_{c}\hat{a}^{\dagger}\hat{a},\nonumber\\
\hat{H}^{(i)}_{I}&=g\left(|e_{i}\rangle\langle g_{i}|\hat{a}+|g_{i}\rangle\langle e_{i}|\hat{a}^{\dagger}\right).
\end{align}

In a previous work, we showed that, assuming all molecules are in their global ground state and taking the large $N$ limit (zeroth-order approximation in CUT-E, single-molecule light matter coupling $g\rightarrow 0$ while keeping $g\sqrt{N}$ constant), the time-dependent wavefunction in the first excitation manifold can be obtained by solving a simple system of equations of motion (EoM) \cite{Juan1}:
\begin{widetext}
\begin{align}
&i\dot{A}^{(0)}_{00...0}(t)=\omega_{c}A^{(0)}_{00...0}(t)+g\sum_{i=1}^{N}\sum_{l_{i}}\langle \varphi_{0}|\phi_{l_{i}}\rangle A^{(i)}_{00...l_{i}...0}(t)\nonumber\\
&i\dot{A}^{(i)}_{00...l_{i}...0}(t)=\left(\omega_{0,i}+\omega^{(i)}_{\nu,l_{i}}\right)A^{(i)}_{00...l_{i}...0}(t)+g\sum_{l_{i}}\langle \phi_{l_{i}}|\varphi_{0}\rangle A^{(0)}_{00...0}(t),
\end{align}
\end{widetext} where $A^{(0)}_{00...0}$ is the amplitude of the state with all molecules in the global ground state and 1 photon in the cavity (superscript 0 indicates cavity) $|1\rangle|\varphi_{0,1}\rangle|\varphi_{0,2}\rangle\cdots |\varphi_{0,N}\rangle$ with frequency $\omega_{c}$, while $A^{(i)}_{00...l_{i}...0}(t)$ is the amplitude of the state where molecule $i$ is excited in the vibrational state $l_{i}$, $|e_{i}\rangle|\varphi_{0,1}\rangle|\varphi_{0,2}\rangle\cdots|\phi_{l_{i},i}\rangle\cdots|\varphi_{0,N}\rangle$, with frequency $\omega_{0,i}+\omega^{(i)}_{\nu,l_{i}}$. Contribution of other states to the total wavefunction vanish at large $N$ \cite{Spano,KeelingZeb,Juan1}.

If the total molecular ensemble consists of $N_{bins}$ groups composed of $N_{i}$ identical molecules each, we can rewrite the EoM as
\begin{widetext}
\begin{align}
&i\dot{A}^{(0)}_{00...0}(t)=\omega_{c}A^{(0)}_{00...0}(t)+g\sum_{i=1}^{N_{bins}}N_{i}\sum_{l_{i}}\langle \varphi_{0,i}|\phi_{l_{i},i}\rangle A^{(i)}_{00...l_{i}...0}(t)\nonumber\\
&i\dot{A}^{(i)}_{00...l_{i}...0}(t)=\left(\omega_{0,i}+\omega^{(i)}_{\nu,l_{i}}\right)A^{(i)}_{00...l_{i}...0}(t)+g\sum_{l_{i}}\langle \phi_{l_{i},i}|\varphi_{0,i}\rangle A^{(0)}_{00...0}(t),
\end{align} where $i$ now runs over the disorder bins and $\sum_{i}N_{i}=N$.
\end{widetext}

By renormalizing the exciton coefficients as $\tilde{A}^{(i)}_{00...l_{i}...0}(t)=\sqrt{N_{i}}A^{(i)}_{00...l_{i}...0}(t)$ we get

\begin{widetext}
\begin{align}\label{eq:EoM1}
&i\dot{A}^{(0)}_{00...0}(t)=\omega_{c}A^{(0)}_{00...0}(t)+g\sqrt{N}\sum_{i=1}^{N_{bins}}\sqrt{P_{i}}\sum_{l_{i}}\langle \varphi_{0,i}|\phi_{l_{i},i}\rangle \tilde{A}^{(i)}_{00...l_{i}...0}(t)\nonumber\\
&i\dot{\tilde{A}}^{(i)}_{00...l_{i}...0}(t)=\left(\omega_{0,i}+\omega^{(i)}_{\nu,l_{i}}\right)\tilde{A}^{(i)}_{00...l_{i}...0}(t)+g\sqrt{N}\sqrt{P_{i}}\sum_{l_{i}}\langle \phi_{l_{i},i}|\varphi_{0,i}\rangle A^{(0)}_{00...0}(t),
\end{align} 
\end{widetext} with $P_{i}=N_{i}/N$. These EoM yield the effective Hamiltonian

\begin{align}\label{eq:ham3si}
\hat{H}^{\prime}_{eff}=&\omega_{c}|1\rangle\langle 1|+\sum_{i}^{N_{bins}}\hat{H}_{e,i}(q_{i})|e_{i}\rangle\langle e_{i}|\nonumber\\
&+g\sqrt{N}\sum_{i}^{N_{bins}}\sqrt{P_{i}}\left(|e_{i}\rangle\langle 1|+|1\rangle\langle e_{i}|\right)\mathds{P}_{i}.
\end{align} The vibrational projector $\mathds{P}_{i}=|\varphi_{0,i}\rangle\langle \varphi_{0,i}|$ implies that the strong light-matter interaction only occurs at the Franck-Condon region. Notice that $\hat{H}^{\prime}_{eff}$ spans the Hilbert space formed by the vibrational and electronic degrees of freedom of all molecules. Equation (\ref{eq:ham3}) in the manuscript is just an extension of this Hamiltonian, acknowledging the product state $|e_{2}\rangle$ and the diabatic coupling between reactant and product $v_{12}$ (see Fig. \ref{fig:PESs}).

In this article, we are only interested in local observables (e.g., excited state populations); hence, we can directly compute them from the effective wave function evolving according to $\hat{H}^{\prime}_{eff}$ [see Ref. \cite{Juan1}, Eq. (23)]. However, $\hat{H}^{\prime}_{eff}$ can be further simplified. This becomes clearer if we assume the Franck-Condon state is identical for each effective molecule, i. e., $|\varphi_{0,i}\rangle\rightarrow|\varphi_{0}\rangle$.
\begin{widetext}
\begin{align}\label{eq:EoM2}
&i\dot{A}^{(0)}_{00...0}(t)=\omega_{c}A^{(0)}_{00...0}(t)+g\sqrt{N}\sum_{i=1}^{N_{bins}}\sqrt{P_{i}}\sum_{l_{i}}\langle \varphi_{0}|\phi_{l_{i},i}\rangle \tilde{A}^{(i)}_{00...l_{i}...0}(t)\nonumber\\
&i\dot{\tilde{A}}^{(i)}_{00...l_{i}...0}(t)=\left(\omega_{0,i}+\omega^{(i)}_{\nu,l_{i}}\right)\tilde{A}^{(i)}_{00...l_{i}...0}(t)+g\sqrt{N}\sqrt{P_{i}}\sum_{l_{i}}\langle \phi_{l_{i},i}|\varphi_{0}\rangle A^{(0)}_{00...0}(t).
\end{align}
\end{widetext} 

One can check that the following effective Hamiltonian and wavefunction leads to the EoM above,
\begin{widetext}
    
\begin{align}
  &\hat{H}_{eff}=\omega_{c}|1\rangle\langle 1|+\sum_{i}^{N_{bins}}\hat{H}_{e,i}(q)|e_{i}\rangle\langle e_{i}|+g\sqrt{N}\sum_{i}^{N_{bins}}\sqrt{P_{i}}\left(|e_{i}\rangle\langle 1|+|1\rangle\langle e_{i}|\right)\mathds{P},\nonumber\\
  &|\tilde{\Psi}(t)\rangle=A^{(0)}_{00...0}(t)|\varphi_{0}\rangle\otimes|1\rangle+\sum_{i}^{N_{bins}}\sum_{l_{i}}\tilde{A}^{(i)}_{00...l_{i}...0}(t)|\phi_{l_{i},i}\rangle\otimes|e_{i}\rangle.
\end{align}

\end{widetext}

This effective Hamiltonian spans the Hilbert space formed by $N_{bins}$ electronic states, but only the vibrational degrees of freedom of a single molecule (see Fig. \ref{fig:linear}). Again, we can directly compute local observables from the effective wave function evolving according to $\hat{H}_{eff}$.  Alternatively, it can be interpreted as pertaining to a single effective molecule with $N_{bins}$ electronic states. This reduction in the electronic and vibrational degrees of freedom is a dramatic simplification compared to the original Hamiltonian in Eq. (\ref{eq:original}). 

The Hamiltonian in Eq. (\ref{eq:ham5}) of the manuscript is an extension of $\hat{H}_{eff}$ for two electronic excited states diabatically coupled and one vibrational mode per molecule.

\setcounter{equation}{0}
\setcounter{figure}{0}

\renewcommand{\thefigure}{S\arabic{figure}}
\renewcommand{\theequation}{S\arabic{equation}}

\section{Convergence analysis}

\begin{figure*}[htb]
\begin{center}
\includegraphics[width=1\linewidth]{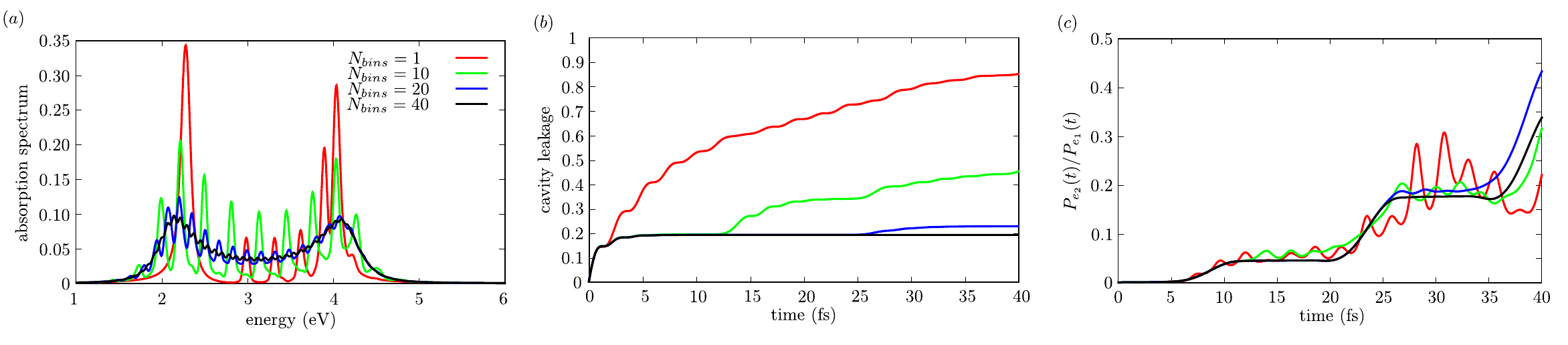}
\caption{Convergence of the dynamics with the number of bins $N_{bins}$. (a) Linear absorption spectrum $A(\omega)$. (b) Cavity leakage $\Gamma(t)$, (c) Ratio of electronic excited state populations $P_{e_{2}}(t)/P_{e_{1}}(t)$. We find that the number of bins $N_{bins}$ needed to reach convergence obeys $N_{bins}=6\sigma\cdot T_{f}/2\pi$. Parameters: $\omega_{0}=0.11$ au, $\sigma=0.02$ au, $\omega_{\nu}=0.01$ au, $g\sqrt{N}=0.03$ au, $\kappa=0.006$ au, $v_{12}=0.0025$ au, $s_{1}=-1$, and $s_{2}=-4$.}
\label{fig:convergence1}
\end{center}
\end{figure*}

In Figure \ref{fig:convergence1} we analyze the convergence in cavity leakage $\Gamma(t)=1-\langle\Psi(t)|\Psi(t)\rangle$, photon population $|C(t)|^{2}$, linear absorption spectrum $A(\omega)$, and electronic populations $P_{e_{1}}(t)$ and $P_{e_{2}}(t)$, as a function of $N_{bins}$. We propagate the initially photonic wavepacket $| \Psi(0) \rangle=|\varphi_{0}\rangle\otimes|1\rangle$ for $40$ fs. 

\section{Effects of cavity leakage in total reactivity: normalized product populations}

\begin{figure}[htb]
\begin{center}
\includegraphics[width=1\linewidth]{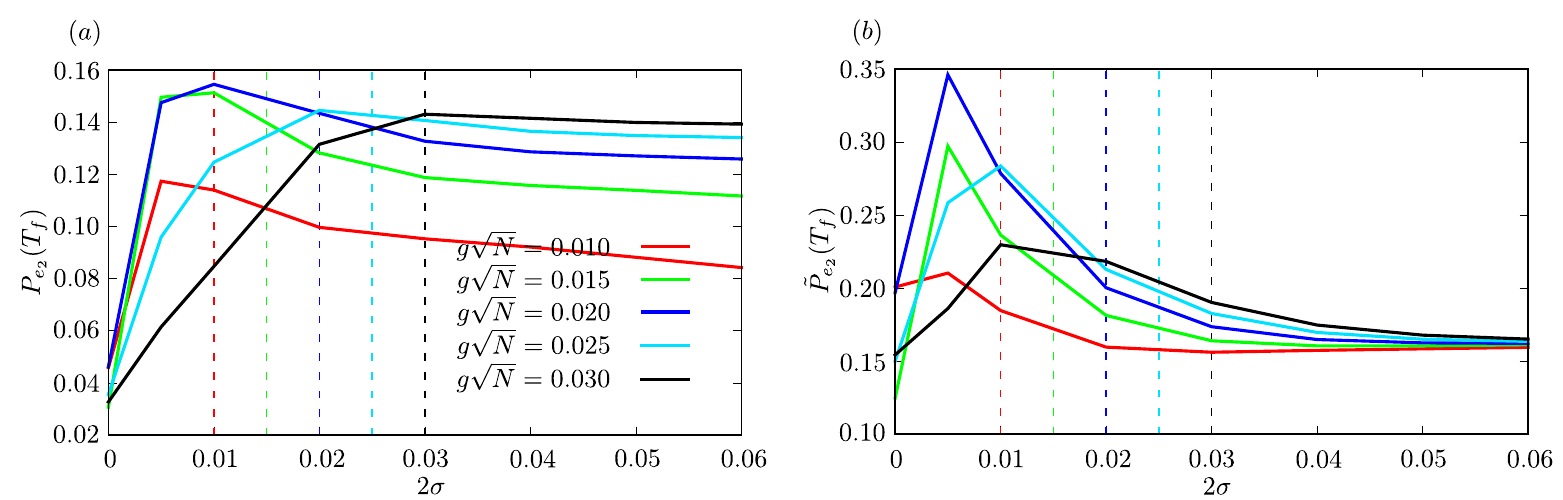}
\caption{Total product population before (a) and after (b) renormalization of the populations. It is clear from this figure that variations at large disorder correspond only to differences in how much energy is lost via cavity leakage.}
\label{fig:cavleak}
\end{center}
\end{figure}

Figure 1c of the main manuscript shows the dependence of the total product population of the ensemble $P_{2}(T_{f})$ at the final time of the simulation ($T_{f}=30$ fs), as a function of disorder and collective light matter coupling strength, after broadband excitation. Looking at the large disorder limit, it seems there is a strong dependence of the reactivity on the coupling strength, opposite to what we claim in the manuscript. In Figure \ref{fig:cavleak} we show that such dependence corresponds simply to differences in cavity leakage. We do this by plotting renormalized populations $\tilde{P}_{2}(T_{f})=P_{2}(T_{f})/\langle\Psi(t)|\Psi(t)\rangle$. 

\section{Detailed analysis of the reactivity of each disorder bin}

\begin{figure*}[htb]
\begin{center}
\includegraphics[width=1\linewidth]{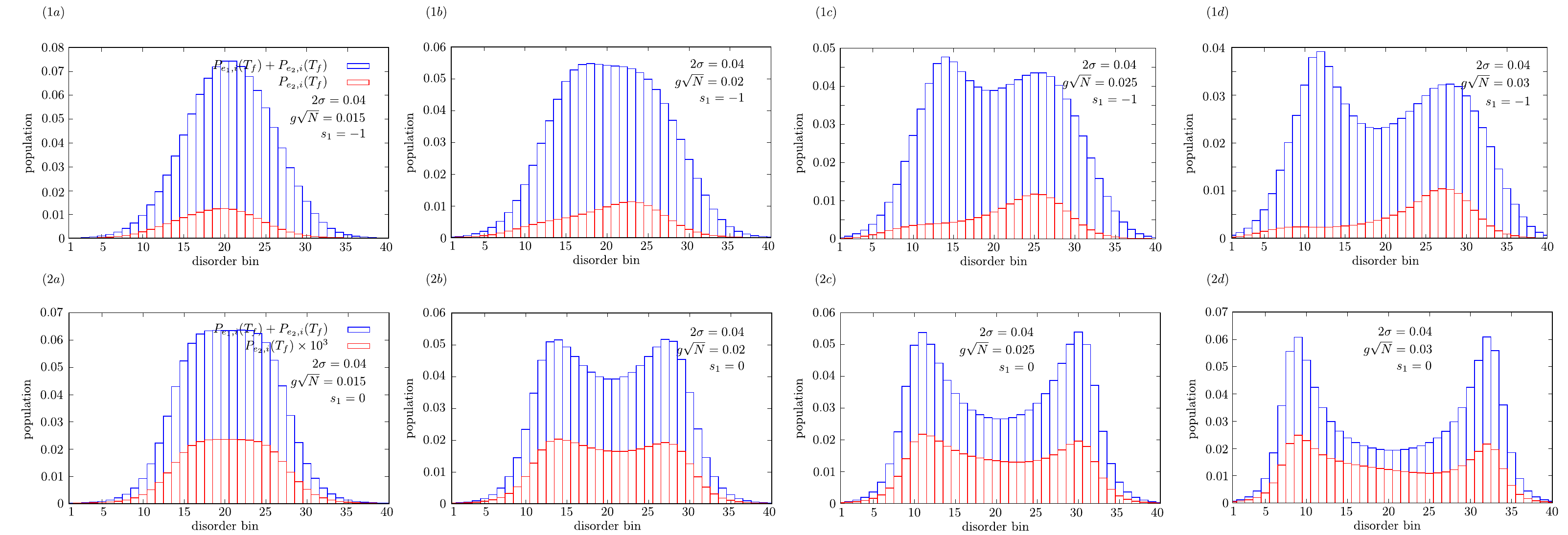}
\caption{Total excited-state population $P_{e_{2,i}}(T_{f})+P_{e_{1,i}}(T_{f})$ and product population $P_{e_{2,i}}(T_{f})$ for each of the disorder bins as a function of collective coupling (keeping disorder fixed). Top and bottom rows show calculations with and without vibronic coupling. Disorder is fixed at $2\sigma=0.04$ au while collective coupling $g\sqrt{N}$ varies from $0-0.03$ au (left to right). In the top row $\omega_{c}$ is resonant with the $\nu=0\rightarrow\nu^{\prime}=1$ transition, while in the bottom row $\omega_{c}$ is resonant with the $\nu=0\rightarrow\nu^{\prime}=0$ transition.}\label{fig:bins}
\end{center}
\end{figure*}

We first calculate the total excited-state populations ($P_{e_{2},i}(T_{f})+P_{e_{1},i}(T_{f})$) as well as  product populations ($P_{e_{2},i}(T_{f})$), for each of the disorder bins. In the top row of Figure \ref{fig:bins} we can see the larger reactivity of high-frequency disorder bins arises only under strong coupling, and cannot be attributed to increase in their respective absorption. In the bottom row we see that this effect is negligible in the absence of vibronic coupling. 

\begin{figure*}[htb]
\begin{center}
\includegraphics[width=1\linewidth]{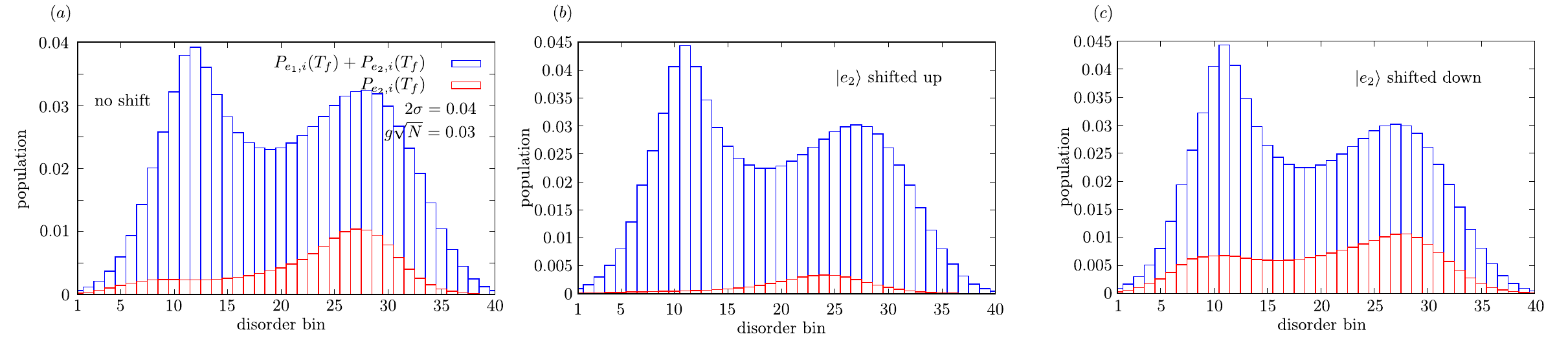}
\caption{total excited state population $P_{e_{2,i}}(T_{f})+P_{e_{1,i}}(T_{f})$ and product population $P_{e_{2,i}}(T_{f})$ for each of the disorder bins after shifting the electronic states $|e_{2,i}\rangle$ up and down by two vibrational quanta ($0.02$ au).}
\label{fig:bins2}
\end{center}
\end{figure*}

Second, we carry out a similar analysis for a slightly modified molecular system, where the $|e_{2}\rangle$ state has been shifted to higher and lower energy, for all disorder bins (see Figure \ref{fig:bins2}). In both cases, disorder bins near the upper-polariton energy still contribute more to the reactivity, and this effect decreases as the barrier for the reaction gets lower. This is consistent with high frequency bins having more kinetic energy in the vibrational degrees of freedom than low-frequency bins. Strong coupling is required so that UP (LP) can target highly (slightly) reactive states. 

\begin{figure}[htb]
\begin{center}
\includegraphics[width=1\linewidth]{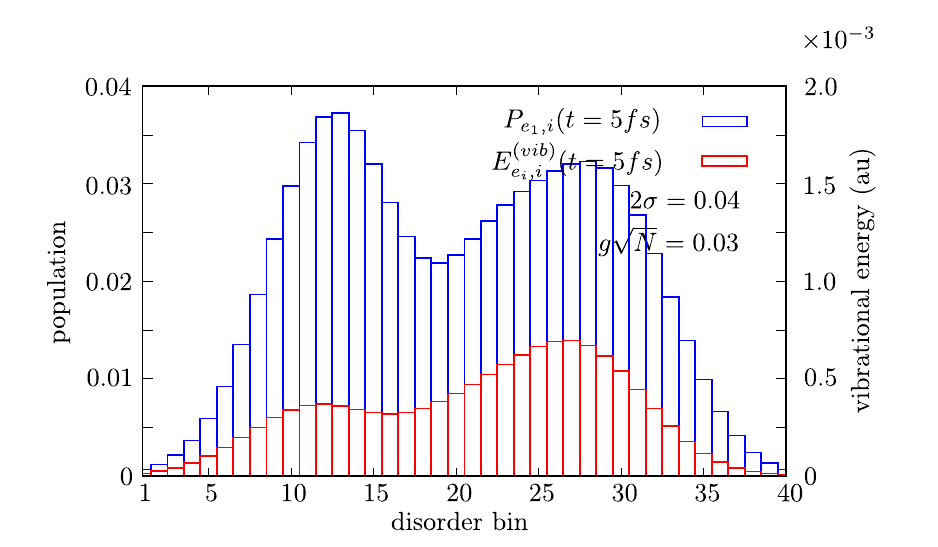}
\caption{Vibrational energy and population of the reactant wavepackets, $E^{(vib)}_{e_{1},i}(t=5$fs$)$ and $P_{e_{1,i}}(t=5$fs$)$, for each of the disorder bins.}
\label{fig:bins4}
\end{center}
\end{figure}

Finally, we provide a further confirmation that higher vibrational energy endowed to higher-frequency bins. We analyze the vibrational wavepackets created in the reactant state $|e_{1,i}\rangle$ of each disorder bin, in the strong coupling regime, and right after damping of the Rabi oscillations ($5$ fs, before the reaction ensues). We calculate the vibrational energy in each bin and compare it with its corresponding excited state population. This reveals that vibrational wavepackets near the upper polariton band acquire more kinetic energy, which explains their higher reactivity (see Figure \ref{fig:bins4}). Future works will establish whether these states can be created outside of the cavity with conventional linear optical sources.

\clearpage
\bibliographystyle{unsrt}
\bibliography{main}

\begin{thebibliography}{10}

\bibitem{Ebbesen}
James~A. Hutchison, Tal Schwartz, Cyriaque Genet, Eloïse Devaux, and Thomas~W.
  Ebbesen.
\newblock Modifying chemical landscapes by coupling to vacuum fields.
\newblock {\em Angew. Chem. Int. Ed.}, 51(7):1592--1596, 2012.

\bibitem{Hongfei}
Hongfei Zeng, Juan~B. Pérez-Sánchez, Christopher~T. Eckdahl, Pufan Liu,
  Woo~Je Chang, Emily~A. Weiss, Julia~A. Kalow, Joel Yuen-Zhou, and
  Nathaniel~P. Stern.
\newblock Control of photoswitching kinetics with strong light–matter
  coupling in a cavity.
\newblock {\em J. Am. Chem. Soc.}, 145(36):19655--19661, 2023.

\bibitem{Blake}
Wonmi Ahn, Johan~F. Triana, Felipe Recabal, Felipe Herrera, and Blake~S.
  Simpkins.
\newblock Modification of ground-state chemical reactivity via light-matter
  coherence in infrared cavities.
\newblock {\em Science}, 380(6650):1165--1168, 2023.

\bibitem{Owrutsky}
Adam~D. Dunkelberger, Blake~S. Simpkins, Igor Vurgaftman, and Jeffrey~C.
  Owrutsky.
\newblock Vibration-cavity polariton chemistry and dynamics.
\newblock {\em Annu. Rev. Phys. Chem.}, 73(1):429--451, 2022.

\bibitem{RaphaelReview}
Raphael~F. Ribeiro, Luis~A. Martínez-Martínez, Matthew Du, Jorge
  Campos-Gonzalez-Angulo, and Joel Yuen-Zhou.
\newblock Polariton chemistry: controlling molecular dynamics with optical
  cavities.
\newblock {\em Chem. Sci.}, 9:6325--6339, 2018.

\bibitem{Wei}
Bo~Xiang, Raphael~F. Ribeiro, Matthew Du, Liying Chen, Zimo Yang, Jiaxi Wang,
  Joel Yuen-Zhou, and Wei Xiong.
\newblock Intermolecular vibrational energy transfer enabled by microcavity
  strong light-matter coupling.
\newblock {\em Science}, 368(6491):665--667, 2020.

\bibitem{TT}
Teng-Teng Chen, Matthew Du, Zimo Yang, Joel Yuen-Zhou, and Wei Xiong.
\newblock Cavity-enabled enhancement of ultrafast intramolecular vibrational
  redistribution over pseudorotation.
\newblock {\em Science}, 378(6621):790--794, 2022.

\bibitem{Jorge2}
J.~A. Campos-Gonzalez-Angulo, Y.~R. Poh, M.~Du, and J.~Yuen-Zhou.
\newblock {Swinging between shine and shadow: Theoretical advances on thermally
  activated vibropolaritonic chemistry}.
\newblock {\em J. Chem. Phys.}, 158(23):230901, 06 2023.

\bibitem{Huo}
Arkajit Mandal and Pengfei Huo.
\newblock Investigating new reactivities enabled by polariton photochemistry.
\newblock {\em J. Phys. Chem. Lett.}, 10(18):5519--5529, 2019.

\bibitem{Herrera}
Felipe Herrera and Jeffrey Owrutsky.
\newblock Molecular polaritons for controlling chemistry with quantum optics.
\newblock {\em J. Chem. Phys.}, 152(10):100902, 2020.

\bibitem{Feist1}
Javier Galego, Francisco~J Garcia-Vidal, and Johannes Feist.
\newblock Suppressing photochemical reactions with quantized light fields.
\newblock {\em Nature Commun.}, 7:13841, 2016.

\bibitem{GalegoRev}
M\'onica S\'anchez-Barquilla, Antonio~I. Fern\'andez-Dom\'inguez, Johannes
  Feist, and Francisco~J. Garc\'ia-Vidal.
\newblock A theoretical perspective on molecular polaritonics.
\newblock {\em ACS Photonics}, 9(6):1830--1841, 2022.

\bibitem{Genes}
Michael Reitz, Christian Sommer, and Claudiu Genes.
\newblock Langevin approach to quantum optics with molecules.
\newblock {\em Phys. Rev. Lett.}, 122:203602, May 2019.

\bibitem{HuoRev}
Arkajit Mandal, Michael Taylor, Braden Weight, Eric Koessler, Xinyang Li, and
  Pengfei Huo.
\newblock Theoretical advances in polariton chemistry and molecular cavity
  quantum electrodynamics.
\newblock {\em ChemRxiv}, 2022.

\bibitem{SKC1}
Tomohiro Ishii, Fatima Bencheikh, Sébastien Forget, Sébastien Chénais,
  Benoît Heinrich, David Kreher, Lydia Sosa~Vargas, Kiyoshi Miyata, Ken Onda,
  Takashi Fujihara, Stéphane Kéna-Cohen, Fabrice Mathevet, and Chihaya
  Adachi.
\newblock Enhanced light–matter interaction and polariton relaxation by the
  control of molecular orientation.
\newblock {\em Adv. Opt. Mater.}, 9(22):2101048, 2021.

\bibitem{SKC2}
Christopher~R. Gubbin, Stefan~A. Maier, and Stéphane Kéna-Cohen.
\newblock Low-voltage polariton electroluminescence from an ultrastrongly
  coupled organic light-emitting diode.
\newblock {\em Appl. Phys. Lett.}, 104(23):233302, 2014.

\bibitem{TaoLi2}
Tao~E. Li and Sharon Hammes-Schiffer.
\newblock Qm/mm modeling of vibrational polariton induced energy transfer and
  chemical dynamics.
\newblock {\em J. Am. Chem. Soc.}, 145(1):377--384, 2023.

\bibitem{Spano2}
F.~C. Spano.
\newblock Optical microcavities enhance the exciton coherence length and
  eliminate vibronic coupling in j-aggregates.
\newblock {\em J. Chem. Phys.}, 142(18):184707, 2015.

\bibitem{Spano}
Frank~C. Spano.
\newblock Exciton–phonon polaritons in organic microcavities: Testing a
  simple ansatz for treating a large number of chromophores.
\newblock {\em J. Chem. Phys.}, 152(20):204113, 2020.

\bibitem{Zhang}
Zhedong Zhang and Shaul Mukamel.
\newblock Fluorescence spectroscopy of vibronic polaritons of molecular
  aggregates in optical microcavities.
\newblock {\em Chem. Phys. Lett.}, 683:653--657, 2017.

\bibitem{ribeiroNat}
Raphael~F. Ribeiro.
\newblock Multimode polariton effects on molecular energy transport and
  spectral fluctuations.
\newblock {\em Commun. Chem.}, 5(1):48, Apr 2022.

\bibitem{Mandal}
Arkajit Mandal, Ding Xu, Ankit Mahajan, Joonho Lee, Milan Delor, and David~R.
  Reichman.
\newblock Microscopic theory of multimode polariton dispersion in multilayered
  materials.
\newblock {\em Nano Letters}, 23(9):4082--4089, 2023.

\bibitem{Cao}
Georg Engelhardt and Jianshu Cao.
\newblock Unusual dynamical properties of disordered polaritons in
  microcavities.
\newblock {\em Phys. Rev. B}, 105:064205, Feb 2022.

\bibitem{Vendrellayer}
Oriol Vendrell and Hans-Dieter Meyer.
\newblock Multilayer multiconfiguration time-dependent hartree method:
  Implementation and applications to a henon–heiles hamiltonian and to
  pyrazine.
\newblock {\em J. Chem. Phys.}, 134(4):044135, 2011.

\bibitem{Neepa2}
Lionel Lacombe, Norah~M. Hoffmann, and Neepa~T. Maitra.
\newblock Exact potential energy surface for molecules in cavities.
\newblock {\em Phys. Rev. Lett.}, 123:083201, Aug 2019.

\bibitem{Tichauer}
Ruth~H. Tichauer, Johannes Feist, and Gerrit Groenhof.
\newblock Multi-scale dynamics simulations of molecular polaritons: The effect
  of multiple cavity modes on polariton relaxation.
\newblock {\em J. Chem. Phys.}, 154(10):104112, 2021.

\bibitem{Groenhof}
Hoi~Ling Luk, Johannes Feist, J.~Jussi Toppari, and Gerrit Groenhof.
\newblock Multiscale molecular dynamics simulations of polaritonic chemistry.
\newblock {\em J. Chem. Theory Comput.}, 13(9):4324--4335, 2017.

\bibitem{Rubio1}
M.~A. Sentef, M.~Ruggenthaler, and A.~Rubio.
\newblock Cavity quantum-electrodynamical polaritonically enhanced
  electron-phonon coupling and its influence on superconductivity.
\newblock {\em Sci. Adv.}, 4(11):eaau6969, 2018.

\bibitem{Rubio2}
Christian Sch\"{a}fer, Michael Ruggenthaler, Heiko Appel, and Angel Rubio.
\newblock Modification of excitation and charge transfer in cavity
  quantum-electrodynamical chemistry.
\newblock {\em PNAS}, 116(11):4883--4892, 2019.

\bibitem{Dominik1}
Dominik Sidler, Christian Schäfer, Michael Ruggenthaler, and Angel Rubio.
\newblock Polaritonic chemistry: Collective strong coupling implies strong
  local modification of chemical properties.
\newblock {\em J. Phys. Chem. Lett.}, 12(1):508--516, 2021.

\bibitem{Sukharev}
Maxim Sukharev, Joseph Subotnik, and Abraham Nitzan.
\newblock {Dissociation slowdown by collective optical response under strong
  coupling conditions}.
\newblock {\em J. Chem. Phys.}, 158(8):084104, 02 2023.

\bibitem{Srihari}
Subhadip Mondal, Derek~S. Wang, and Srihari Keshavamurthy.
\newblock {Dissociation dynamics of a diatomic molecule in an optical cavity}.
\newblock {\em J. Chem. Phys.}, 157(24):244109, 12 2022.

\bibitem{Houdre}
R.~Houdr\'e, R.~P. Stanley, and M.~Ilegems.
\newblock Vacuum-field rabi splitting in the presence of inhomogeneous
  broadening: Resolution of a homogeneous linewidth in an inhomogeneously
  broadened system.
\newblock {\em Phys. Rev. A}, 53:2711--2715, Apr 1996.

\bibitem{Schachenmayer3}
David Wellnitz, Guido Pupillo, and Johannes Schachenmayer.
\newblock Disorder enhanced vibrational entanglement and dynamics in
  polaritonic chemistry.
\newblock {\em Commun. Phys.}, 5(1):120, May 2022.

\bibitem{Borrelli}
Maxim~F. Gelin, Amalia Velardo, and Raffaele Borrelli.
\newblock Efficient quantum dynamics simulations of complex molecular systems:
  A unified treatment of dynamic and static disorder.
\newblock {\em J. Chem. Phys.}, 155(13):134102, 2021.

\bibitem{Zhao}
Kewei Sun, Cunzhi Dou, Maxim~F. Gelin, and Yang Zhao.
\newblock Dynamics of disordered tavis–cummings and
  holstein–tavis–cummings models.
\newblock {\em J. Chem. Phys.}, 156(2):024102, 2022.

\bibitem{Cohn}
Bar Cohn, Shmuel Sufrin, Arghyadeep Basu, and Lev Chuntonov.
\newblock Vibrational polaritons in disordered molecular ensembles.
\newblock {\em J. Phys. Chem. Lett.}, 13(35):8369--8375, 2022.

\bibitem{Rury}
Aleksandr~G. Avramenko and Aaron~S. Rury.
\newblock Light emission from vibronic polaritons in coupled
  metalloporphyrin-multimode cavity systems.
\newblock {\em J. Phys. Chem. Lett.}, 13(18):4036--4045, 2022.

\bibitem{Belyanin}
Mikhail Tokman, Alex Behne, Brandon Torres, Maria Erukhimova, Yongrui Wang, and
  Alexey Belyanin.
\newblock Dissipation-driven formation of entangled dark states in strongly
  coupled inhomogeneous many-qubit systems in solid-state nanocavities.
\newblock {\em Phys. Rev. A}, 107:013721, Jan 2023.

\bibitem{Cao1}
Georg Engelhardt and Jianshu Cao.
\newblock Polariton localization and dispersion properties of disordered
  quantum emitters in multimode microcavities.
\newblock {\em Phys. Rev. Lett.}, 130:213602, May 2023.

\bibitem{Beljonne}
Antonios~M. Alvertis, Raj Pandya, Claudio Quarti, Laurent Legrand, Thierry
  Barisien, Bartomeu Monserrat, Andrew~J. Musser, Akshay Rao, Alex~W. Chin, and
  David Beljonne.
\newblock {First principles modeling of exciton-polaritons in polydiacetylene
  chains}.
\newblock {\em J. Chem. Phys.}, 153(8):084103, 08 2020.

\bibitem{Fassioli}
Francesca Fassioli, Kyu~Hyung Park, Sarah~E. Bard, and Gregory~D. Scholes.
\newblock Femtosecond photophysics of molecular polaritons.
\newblock {\em J. Phys. Chem. Lett.}, 12(46):11444--11459, 2021.

\bibitem{Xu}
Ding Xu, Arkajit Mandal, James~M. Baxter, Shan-Wen Cheng, Inki Lee, Haowen Su,
  Song Liu, David~R. Reichman, and Milan Delor.
\newblock Ultrafast imaging of polariton propagation and interactions.
\newblock {\em Nat. Commun.}, 14(1):3881, Jun 2023.

\bibitem{suyabatmaz}
Enes Suyabatmaz and Raphael~F. Ribeiro.
\newblock {Vibrational polariton transport in disordered media}.
\newblock {\em J. Chem. Phys.}, 159(3):034701, 07 2023.

\bibitem{allard}
Thomas~F. Allard and Guillaume Weick.
\newblock Disorder-enhanced transport in a chain of lossy dipoles strongly
  coupled to cavity photons.
\newblock {\em Phys. Rev. B}, 106:245424, Dec 2022.

\bibitem{Musser}
Raj Pandya, Arjun Ashoka, Kyriacos Georgiou, Jooyoung Sung, Rahul Jayaprakash,
  Scott Renken, Lizhi Gai, Zhen Shen, Akshay Rao, and Andrew~J. Musser.
\newblock Tuning the coherent propagation of organic exciton-polaritons through
  dark state delocalization.
\newblock {\em Adv. Sci.}, 9(18):2105569, 2022.

\bibitem{Chen}
Hsing-Ta Chen, Zeyu Zhou, Maxim Sukharev, Joseph~E. Subotnik, and Abraham
  Nitzan.
\newblock Interplay between disorder and collective coherent response:
  Superradiance and spectral motional narrowing in the time domain.
\newblock {\em Phys. Rev. A}, 106:053703, Nov 2022.

\bibitem{Giebink}
Nina Krainova, Alex~J. Grede, Demetra Tsokkou, Natalie Banerji, and Noel~C.
  Giebink.
\newblock Polaron photoconductivity in the weak and strong light-matter
  coupling regime.
\newblock {\em Phys. Rev. Lett.}, 124:177401, Apr 2020.

\bibitem{Juan1}
Juan~B. P\'erez-S\'anchez, Arghadip Koner, Nathaniel~P. Stern, and Joel
  Yuen-Zhou.
\newblock Simulating molecular polaritons in the collective regime using
  few-molecule models.
\newblock {\em PNAS}, 120(15):e2219223120, 2023.

\bibitem{KeelingZeb}
M.~Ahsan Zeb, Peter~G. Kirton, and Jonathan Keeling.
\newblock Exact states and spectra of vibrationally dressed polaritons.
\newblock {\em ACS Photonics}, 5(1):249--257, 2018.

\bibitem{KeelingZeb2}
M.~Ahsan Zeb.
\newblock Efficient linear scaling mapping for permutation symmetric fock
  spaces.
\newblock {\em Comp. Phys. Commun.}, 276:108347, 2022.

\bibitem{KeelingZeb3}
M.~Ahsan Zeb, Peter~G. Kirton, and Jonathan Keeling.
\newblock Incoherent charge transport in an organic polariton condensate.
\newblock {\em Phys. Rev. B}, 106:195109, Nov 2022.

\bibitem{Cui}
Bingyu Cui and Abraham Nizan.
\newblock {Collective response in light–matter interactions: The interplay
  between strong coupling and local dynamics}.
\newblock {\em J. Chem. Phys.}, 157(11):114108, 09 2022.

\bibitem{Bing}
Bing Gu.
\newblock Toward collective chemistry by strong light-matter coupling.
\newblock {\em arXiv:2306.08944}, 2023.

\bibitem{Dvira}
Nicholas Anto-Sztrikacs and Dvira Segal.
\newblock Capturing non-markovian dynamics with the reaction coordinate method.
\newblock {\em Phys. Rev. A}, 104:052617, Nov 2021.

\bibitem{CederbaumPRL}
Lorenz~S. Cederbaum, Etienne Gindensperger, and Irene Burghardt.
\newblock Short-time dynamics through conical intersections in macrosystems.
\newblock {\em Phys. Rev. Lett.}, 94:113003, Mar 2005.

\bibitem{Cwik}
Justyna~A. Cwik, Peter Kirton, Simone De~Liberato, and Jonathan Keeling.
\newblock Excitonic spectral features in strongly coupled organic polaritons.
\newblock {\em Phys. Rev. A}, 93:033840, Mar 2016.

\bibitem{HerreraPRA}
Felipe Herrera and Frank~C. Spano.
\newblock Absorption and photoluminescence in organic cavity qed.
\newblock {\em Phys. Rev. A}, 95:053867, May 2017.

\bibitem{Arghadip}
Joel Yuen-Zhou and Arghadip Koner.
\newblock Linear response of molecular polaritons.
\newblock {\em ArXiv:2310.15424}, 2023.

\bibitem{Heller}
Eric~J. Heller.
\newblock The semiclassical way to molecular spectroscopy.
\newblock {\em Acc. Chem. Res.}, 14(12):368--375, 1981.

\bibitem{Tannor}
D.J. Tannor.
\newblock {\em Introduction to Quantum Mechanics: A Time Dependent
  Perspective}.
\newblock University Science Books, Melville, 2007.

\bibitem{Gera}
Tarun Gera and K.~L. Sebastian.
\newblock Effects of disorder on polaritonic and dark states in a cavity using
  the disordered tavis–cummings model.
\newblock {\em J. Chem. Phys.}, 156(19):194304, 05 2022.

\bibitem{HerreraPRL}
Felipe Herrera and Frank~C. Spano.
\newblock Cavity-controlled chemistry in molecular ensembles.
\newblock {\em Phys. Rev. Lett.}, 116:238301, Jun 2016.

\bibitem{Kuttruff}
Joel Kuttruff, Marco Romanelli, Esteban Pedrueza-Villalmanzo, Jonas Allerbeck,
  Jacopo Fregoni, Valeria Saavedra-Becerril, Joakim Andr{\'e}asson, Daniele
  Brida, Alexandre Dmitriev, Stefano Corni, and Nicol{\`o} Maccaferri.
\newblock Sub-picosecond collapse of molecular polaritons to pure molecular
  transition in plasmonic photoswitch-nanoantennas.
\newblock {\em Nat. Commun.}, 14(1):3875, Jul 2023.

\bibitem{thomas}
Philip~A. Thomas, Wai~Jue Tan, Vasyl~G. Kravets, Alexander~N. Grigorenko, and
  William~L. Barnes.
\newblock Non-polaritonic effects in cavity-modified photochemistry.
\newblock {\em Adv. Mater.}, 36(7):2309393, 2024.

\bibitem{Groenhoff}
Arpan Dutta, Ville Tiainen, Luis Duarte, Nemanja Markesevic, Dmitry Morozov,
  Hassan~A. Qureshi, Gerrit Groenhof, and J.~Jussi Toppari, 2023.

\end{thebibliography}

\end{document}